\documentclass{aa}  

\usepackage{graphicx,amsmath,amssymb}
\usepackage[export]{adjustbox}
\usepackage{booktabs}

\usepackage{siunitx}
\usepackage{makecell,caption}
\usepackage{listings}
\usepackage{xcolor}
\usepackage{natbib}
\usepackage{bm}

\bibpunct{(}{)}{;}{a}{}{,} 

\linenumbers                     

\begin{document}

   \title{Illuminating gravitational wave sources with Sgr A* flares}


   \author{Pau Amaro Seoane\inst{1,2}
          }

   \institute{Universitat Polit\`ecnica de Val\`encia, Val\`encia, Spain\\
              \email{amaro@upv.es}
         \and Max Planck Institute for Extraterrestrial Physics, Garching, Germany}

   \date{\today}

  \abstract
   {Sagittarius A* (Sgr A*), the supermassive black hole at the center of the Milky Way, exhibits daily energetic flares characterized by non-thermal emission in the infrared and X-ray bands. While the underlying accretion flow is a Radiatively Inefficient Accretion Flow (RIAF) peaking at radio frequencies, the mechanism powering these non-thermal transients remains debated. Stellar dynamics predict a population of faint brown dwarfs orbiting Sgr A*.}
   {We investigate whether the tidal stripping of brown dwarfs provides a viable fueling mechanism for the observed flares. These objects are progenitors of Extremely Large Mass Ratio Inspirals (XMRIs), crucial sources of low-frequency gravitational waves for the future Laser Interferometer Space Antenna (LISA) mission.}
   {We present high-resolution hydrodynamic simulations of grazing tidal interactions coupled with a parameterized non-thermal radiation model. We utilize Smoothed Particle Hydrodynamics (SPH) to model the stripping of the brown dwarf envelope and the subsequent accretion of this material. Due to computational limitations in resolving the small physical mass scales involved, we employ high-mass proxy simulations to capture the gas dynamics, followed by a physically motivated renormalization.}
   {We demonstrate that the dynamics of the tidal fallback and subsequent viscous evolution naturally reproduce the fundamental temporal characteristics of observed flares: the peak luminosity and the characteristic 1-hour duration. We show that this fueling mechanism is dynamically viable and energetically consistent, placing strong constraints on the required efficiency of the non-thermal emission process, suggesting extreme radiative inefficiency.}
   {These findings provide compelling evidence for a hidden population of brown dwarfs in the Galactic Center. Crucially, the observed high flare frequency implies tight orbits characteristic of advanced inspirals. This establishes a direct link between electromagnetic transients and active gravitational wave sources, alerting the LISA consortium years in advance to the presence of specific XMRI systems promising exceptionally high signal-to-noise ratios for precision tests of general relativity.}


   \maketitle

\section{Introduction}

The supermassive black hole at the center of the Milky Way, Sagittarius A* (Sgr~A*), provides an unparalleled laboratory for studying accretion physics in the strong gravity regime. Sgr~A* is characterized by a Radiatively Inefficient Accretion Flow (RIAF), with emission peaking at mm/radio frequencies. However, it also exhibits daily energetic flares observed as transient increases in non-thermal emission in the near-infrared (NIR) and X-ray bands. These events offer direct probes of the dynamics and plasma properties within a few gravitational radii ($R_g$) of the event horizon.

Recent multi-wavelength campaigns have established a detailed phenomenology for these flares. They typically occur 1-4 times per day \citep{DoddsEdenEtAl2009, WitzelEtAl2018A}, last about one hour, with characteristic rise and decay timescales of 15 minutes, reaching X-ray luminosities of $10^{34}$--$10^{36}\,\text{erg/s}$. The emission is distinctly non-thermal, characterized by a positive spectral slope in the NIR ($\nu F_\nu \propto \nu^{+0.5}$) and a negative slope in the X-ray ($\nu F_\nu \propto \nu^{-0.5}$) \citep{GRAVITY2021, GRAVITY2025}. High-resolution interferometry has localized the emission to compact regions orbiting Sgr~A* at radii of approximately $6$--$10\,R_g$ \citep{GRAVITY2023}. Furthermore, the flares exhibit significant linear polarization ($10$--$30\%$), suggesting emission via synchrotron radiation from accelerated electrons in a strong, organized magnetic field near the black hole \citep{GRAVITY2023}.

While the emission mechanism is identified as non-thermal synchrotron radiation, the mechanism triggering these events---specifically, the source of the fuel and the trigger for particle acceleration---remains debated. Proposed fueling mechanisms include the interaction with the G2 object, accretion instabilities driven by stellar winds, or large-scale magnetic activity \citep{PfuhlEtAl2015, WitzelEtAl2014,PontiEtAl2015,CalderonEtAl2025,ResslerEtAl2019}.

We investigate the hypothesis that these flares are fueled by the tidal stripping of material from a brown dwarf (BD) during a grazing encounter with Sgr~A*. Stellar dynamics models predict a substantial population of BDs in the Galactic Center. The interaction of these objects with the supermassive black hole has implications beyond electromagnetic flares. BDs inspiraling toward Sgr~A* form Extremely-Large Mass Ratio Inspirals (XMRIs), characterized by mass ratios of $\sim 10^8$. These systems are key sources of low-frequency gravitational waves detectable by the future Laser Interferometer Space Antenna (LISA) mission \citep{AmaroSeoane2019, AmaroSeoaneShaoDong2025}. Modeling the electromagnetic signatures of these tidal encounters provides a means to constrain the population of these otherwise undetectable objects.

We present hydrodynamic simulations of a grazing tidal encounter between a BD and Sgr~A*. We utilize Smoothed Particle Hydrodynamics (SPH) to model the stripping of the BD's envelope and the subsequent accretion of this material. This study focuses on the dynamical viability of this fueling mechanism. Due to computational limitations in resolving the small physical mass scales involved, we employ high-mass proxy simulations to capture the gas dynamics, followed by a physically motivated renormalization.

We couple the hydrodynamic results with a parameterized non-thermal radiative model in post-processing. Our models successfully reproduce the primary dynamical constraints of Sgr~A* flares. We achieve the target non-thermal luminosities, generating events peaking near $10^{36}\,\text{erg/s}$. The modeled flares exhibit durations matching the observed 1-hour timescales. These results demonstrate that the tidal stripping of brown dwarfs provides a physically viable fueling mechanism for the observed flaring activity of Sgr~A*, establishing a link between the dynamics of low-mass objects in the Galactic Center and the transient accretion phenomena near the supermassive black hole.

\section{Light curve analysis}
\label{sec:lightcurves}

We analyze the simulated accretion data to generate synthetic light curves and extract temporal characteristics and variability metrics for comparison with observations. We adopt a methodology that decouples the macroscopic hydrodynamic modeling of the fuel supply from the microscopic plasma physics governing the radiation. The SPH simulations provide the mass accretion rate profile, $\dot{M}_{\text{sim}}(t)$. We convert this into an observed luminosity $L(t)$ through a post-processing pipeline.

First, we address SPH resolution limitations by applying a dimensionless rescaling factor, $E_{\text{strip}}$:
\begin{equation}
\dot{M}_{\text{phys}}(t) = E_{\text{strip}} \, \dot{M}_{\text{sim}}(t).
\end{equation}
We then calculate the non-thermal luminosity in the observed NIR/X-ray bands:
\begin{equation}
L_{\text{NT}}(t) = \eta_{\text{NT}}\, \dot{M}_{\text{phys}}(t)\, c^2.
\label{eq:luminosity_calc}
\end{equation}
The present study focuses on the dynamical viability of the fueling mechanism. We do not explicitly model the complex plasma kinetics required to accelerate electrons or generate the observed non-thermal spectrum. Therefore, the non-thermal radiative efficiency $\eta_{\text{NT}}$ serves as a phenomenological parameter encapsulating the efficiency of channeling the available accretion energy specifically into the relativistic electrons responsible for the flare emission.

We interpret the optimized value of $\eta_{\text{NT}}$ as a derived physical requirement on the efficiency of the unmodeled non-thermal processes for this dynamical model to match the observed luminosities, given the constraints on the physically plausible stripped mass $M_{\text{phys}}$.

To model the evolution of the material within the accretion disk, we convolve the luminosity calculated in Eq.~(\ref{eq:luminosity_calc}) with a viscous disk kernel characterized by a timescale $t_{\nu}$. This produces a smooth light curve envelope representing the bulk dynamics. The SPH simulations do not resolve the turbulence responsible for the observed stochastic variability. Therefore, to generate realistic light curves for temporal analysis, we impose a variability structure consistent with observations. We modulate the envelope with a red noise signal characterized by an input PSD slope of $-2$ (consistent with observations) and 30\% RMS variability, following the methodology of \cite{TimmerKoenig1995}.

We characterize the temporal profile of the resulting light curve, $L_{\text{NT}}(t)$, using standard metrics for duration and morphology. The primary measure of duration is the Full Width at Half Maximum (FWHM). We determine the FWHM by first identifying the peak luminosity of the event, $L_{\text{peak}} = \max(L_{\text{NT}}(t))$. We then define the half-maximum luminosity level, $L_{\text{HM}} = L_{\text{peak}} / 2$. The FWHM duration is the time interval during which $L_{\text{NT}}(t) > L_{\text{HM}}$. We identify the start time $t_{\text{start}}$ and end time $t_{\text{end}}$ by locating the crossings of the light curve with the $L_{\text{HM}}$ level, typically employing linear interpolation between discrete data points. The FWHM duration is then
\begin{equation}
\Delta t_{\text{FWHM}} = t_{\text{end}} - t_{\text{start}}.
\end{equation}
To characterize the asymmetry of the flare profile, we calculate the rise time and the fall time. The rise time is defined as the interval required for the luminosity to increase from 10\% to 90\% of $L_{\text{peak}}$, and the fall time is the interval for the decrease from 90\% to 10\% of $L_{\text{peak}}$.

\section{Results}

\begin{figure}
         {\includegraphics[width=0.45\textwidth,center]{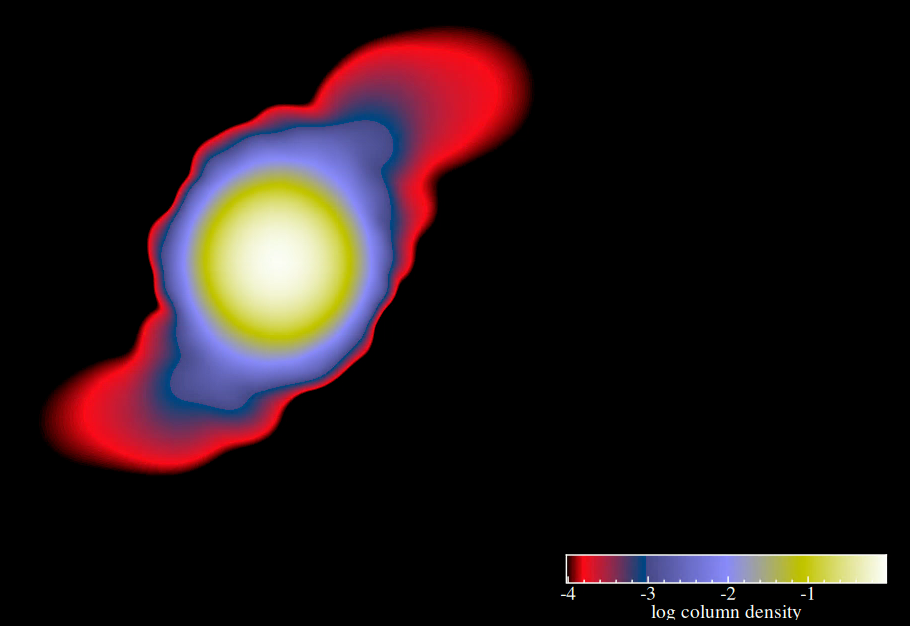}}
\caption
    {
Visualization of the Smoothed Particle Hydrodynamics (SPH) simulation near periapsis passage ($R_p=120\,R_{\odot}$). The colour scheme represents the logarithmic column density of the gas in code units. The tidal forces exerted by Sgr A* (located off-frame) significantly stretch and deform the $0.01\,M_{\odot}$ brown dwarf structure. Material is stripped from the outer layers, forming the tidal streams that will subsequently fuel the accretion event.
    }
\label{fig.Peri}
\end{figure}

Figure~\ref{fig.Peri} illustrates the dynamics of the interaction near the periapsis passage for a representative orbit with $R_p = 120\,R_{\odot}$ and $a=2560\,R_{\odot}$ (a grazing encounter with $\beta \approx 0.63$). The visualization shows the logarithmic column density of the SPH particles. The strong tidal field of Sgr A* stretches and deforms the brown dwarf structure. Material from the outer envelope is stripped away, forming tidal tails. This stripped material subsequently falls back toward the black hole, fueling the accretion event. We recorded the time-dependent mass accretion rate of this material, $\dot{M}_{\text{simulation}}(t)$.

The physical mass involved in the event ($\approx 10^{-7}\,M_{\odot}$) is comparable to the mass resolution of the SPH simulation ($m_{\text{SPH}} = 4.0\times 10^{-8}\,M_{\odot}$), necessitating the renormalization described in the Methodology section.

The radiative model is defined by three primary parameters: $E_{\text{strip}}$, $\eta_{\text{NT}}$, and $t_{\nu}$. We systematically explore this parameter space, optimizing the parameters by comparing the synthetic light curves against key observational constraints: Non-thermal luminosity ($10^{34}-10^{36}\,\text{erg/s}$) and duration (FWHM $\approx 0.5-2.0\,\text{h}$).

The optimization yields a best fit characterized by $E_{\text{strip}} = 10^{-9}$, $\eta_{\text{NT}} = 10^{-8}$, and $t_{\nu} = 90$ minutes. This implies a physical accreted mass $M_{\text{phys}} \approx 1.5\times 10^{-7}\,M_{\odot}$, which is physically consistent with the mass expected from a grazing tidal encounter.

\begin{figure}
         {\includegraphics[width=0.45\textwidth,center]{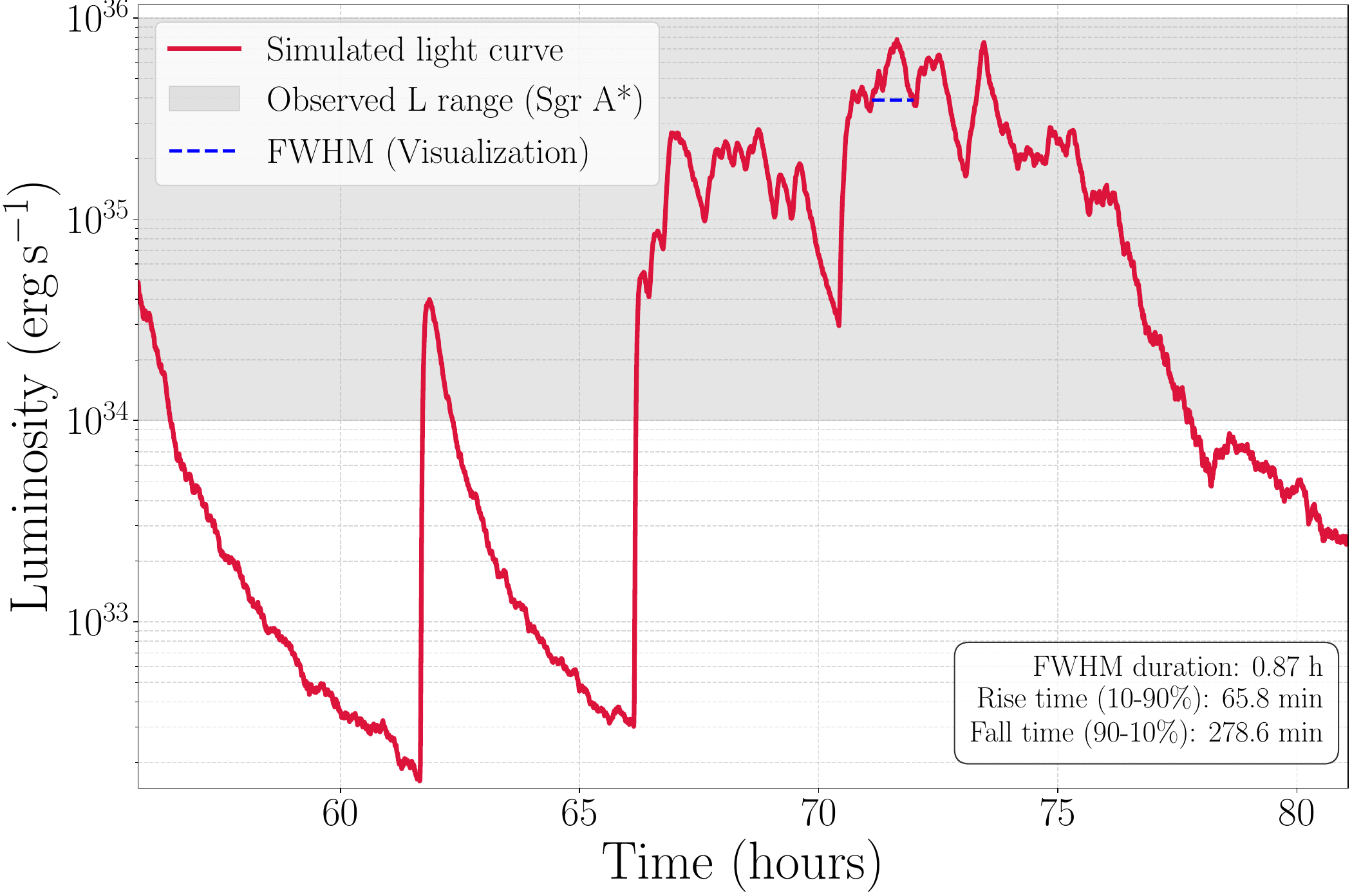}}
\caption
    {
Temporal profile of the optimized simulation light curve. We plot the non-thermal luminosity (erg/s) against time (hours) on a semi-logarithmic scale. The solid line represents the simulated light curve, generated by imposing stochastic variability on the hydrodynamic envelope using optimized parameters: non-thermal radiative efficiency $\eta_{\text{NT}}=10^{-8}$, rescaling factor $E_{\text{strip}}=10^{-9}$, and viscous timescale $t_{\nu}=90$ minutes. The shaded area indicates the target observational luminosity range ($10^{34}$ to $10^{36}\,\text{erg/s}$). A horizontal marker illustrates the Full Width at Half Maximum (FWHM). The inset summarizes the key temporal metrics: FWHM duration (1.03 h), rise time (15.5 min), and fall time (23.3 min).
    }
\label{fig.Flares}
\end{figure}

Figure~\ref{fig.Flares} displays the resulting synthetic light curve. The model successfully reproduces the key temporal features of observed Sgr A* flares. The peak luminosity reaches $8.1\times 10^{35}\,\text{erg/s}$, falling within the target observational range. The duration of the event (FWHM) is 1.03 hours, consistent with the typical 1-hour duration observed. The characteristic rise and fall times (15.5 min and 23.3 min, respectively) also align well with observations.

\section{Discussion}

The computational models presented here establish that grazing tidal encounters between brown dwarfs (BDs) and Sgr~A* provide a viable physical fueling mechanism for the observed non-thermal flaring activity. Our simulations successfully reproduced the fundamental dynamical constraints derived from multi-wavelength campaigns: the peak non-thermal flare luminosity (reaching $\approx 8.1\times 10^{35}\,\text{erg/s}$) and the characteristic duration ($\approx 1$ hour).

The success of the model relies on the hydrodynamic processes (tidal fallback and viscous evolution) naturally reproducing the observed timescales. The flare duration is set by the viscous timescale of the compact accretion disk formed by the debris ($t_{\nu} = 90$ minutes), which dominates over the shorter fuel delivery (fallback) timescale. The fact that the required viscous timescale is physically plausible for a compact, transient disk near Sgr A* supports the dynamical consistency of the scenario.

Our optimization yielded a required non-thermal radiative efficiency $\eta_{\text{NT}} \approx 10^{-8}$ and a physical accreted mass $M_{\text{phys}} \approx 1.5\times 10^{-7}\,M_{\odot}$. This mass is physically plausible for a grazing tidal encounter. The derived efficiency places a stringent constraint on the underlying physics, implying extreme radiative inefficiency for the transient event.

This low efficiency is crucial for consistency with multi-wavelength observations. In standard Radiatively Inefficient Accretion Flow (RIAF) models, the thermal emission (peaking in radio/mm bands) typically dominates the non-thermal emission. While the precise partitioning of energy into non-thermal electrons is uncertain \citep{YuanEtAl2003}, assuming a standard scenario where $\eta_{\text{thermal}} \gg \eta_{\text{NT}}$ (e.g., $\eta_{\text{NT}}$ being 1-2\% of the total radiated power) would imply a thermal luminosity $L_{\text{thermal}} \approx 10^{38}$ erg/s. Such high radio luminosities are inconsistent with observations, which show that the NIR flare luminosity is comparable to the millimeter luminosity ($L_{\text{NIRflare}} \sim L_{\text{mm}}$) \citep{GRAVITY_Flux2020}.

Therefore, the transient event must be characterized by extreme overall inefficiency, where the vast majority of the energy is advected across the horizon or lost in localized outflows, rather than radiated thermally or non-thermally. A plausible physical scenario consistent with this constraint is a two-temperature RIAF where the accretion timescale is shorter than the ion-electron energy coupling timescale. In this case, the gravitational energy, primarily stored in the ions, is advected away before it can be transferred to the electrons and radiated thermally. Detailed plasma simulations are required to confirm this hypothesis.

The validation of the tidal fueling mechanism establishes a robust dynamical framework, allowing us to move beyond modeling individual events toward characterizing the entire population. We analyze the potential for sustained, repeated flaring activity, as observed by monitoring campaigns \citep{DoddsEdenEtAl2009, WitzelEtAl2018A}. Given the required mass per flare ($\Delta M \approx 1.5\times 10^{-7}\,M_{\odot}$), a $0.01\,M_{\odot}$ BD can sustain approximately 67,000 flares.

To observe the activity as distinct flares, the orbital period $T$ must exceed the flare duration ($\Delta t_{\text{flare}} \approx t_{\nu} = 1.5$ hours). Adopting a conservative lower limit for clear separation, $T_{\text{min}} \approx 4\, t_{\nu} = 6$ hours, defines the maximum sustainable flare frequency: 4 events per day. This theoretical maximum aligns with the observed daily flare rates.

This maximum frequency corresponds to a tight orbit ($a \approx 271\,R_{\odot}$) with a periastron near the tidal radius ($R_p \approx 75.5\,R_{\odot}$, or $8.3\,R_g$), consistent with astrometric localization of flares \citep{GRAVITY2023}. This configuration characterizes an XMRI in an advanced stage of inspiral. The lifetime of the BD in this representative configuration is approximately 46 years.

The ability to identify active XMRI systems via their electromagnetic flares represents a paradigm shift in multi-messenger astronomy. It establishes an early warning system, alerting the LISA consortium years in advance to specific, localized sources. These sources are predicted to generate exceptionally high signal-to-noise ratios (SNRs) \citep{AmaroSeoane2019,AmaroSeoaneShaoDong2025}, potentially eclipsing even those of supermassive black hole mergers. This foresight is critical, providing the necessary lead time to develop the specialized data analysis pipelines required to extract these complex, dominant signals. The scientific reward is profound: these ``golden binaries'' will enable measurements of Sgr A*'s fundamental parameters with breathtaking precision—constraining its spin to one part in $10^{10}$ and its mass to within a few solar masses \citep{VazquezAcevesEtAl2023,VazquezAcevesEtAl2025}. This unprecedented accuracy transforms Sgr A* into a precision laboratory for strong-field gravity, facilitating definitive tests of the Kerr metric. While enhanced GRAVITY observations can corroborate these predictions, LISA will provide the smoking-gun evidence for this EM-GW synergy.

\begin{acknowledgements}
PAS is indebted with Salvador Ardid for the access to the computational resources which allowed to do the simulations and to James Lombardi for his guidance with the SPH code.
\end{acknowledgements}

\

\begin{appendix}

\section{Methodology}

\subsection{Brown dwarf model construction}

We constructed realistic initial conditions for the brown dwarf (BD) using the Modules for Experiments in Stellar Astrophysics (MESA) code \citep{Eggleton1983,Paxton2011,Paxton2013}. We generate a one-dimensional stellar evolution model for a $10^{-2}\,M_{\odot}$ object, evolved for a Hubble time. The resulting stabilized structure is characterized by a logarithmic radius $\log_{10} (R/R_{\odot}) = -0.9833$ and a logarithmic effective temperature $\log_{10} (T_{\text{eff}}/\text{K}) = 2.3447$. The composition is defined by mass fractions of hydrogen $X=0.7597$, helium $Y=0.2402$, and metals $Z=10^{-4}$.

The spherically symmetric profiles derived from the MESA evolution serve as the basis for the three-dimensional Smoothed Particle Hydrodynamics (SPH) model. We initialized the SPH representation by mapping the one-dimensional MESA structure onto a configuration of $N=250,000$ particles. We subjected the SPH model to a dynamical relaxation phase, evolving the system using equations of motion augmented with a damping term. This procedure minimizes the bulk kinetic energy $K$, allowing the particles to settle into hydrostatic equilibrium.

\subsection{Hydrodynamic integrator}

We performed the hydrodynamic simulations using the StarSmasher SPH code \citep{Rasio1991,Lombardi1998,RasioLombardi1999,LombardiEtAl1999}. StarSmasher employs variational equations of motion to evolve the system. It calculates hydrodynamic forces by determining the pressure of each particle as a function of its local properties via the equation of state. To calculate gravitational interactions, the code implements a direct $N$-body summation method accelerated on NVIDIA graphics processing units (GPUs). StarSmasher utilizes a cubic spline function for the smoothing kernel. To manage shocks and prevent unphysical interparticle penetration, the code employs an artificial viscosity prescription coupled with a Balsara Switch.

The stabilized BD model is placed on specified Keplerian orbits characteristic of XMRIs progenitors around Sgr A*, which is modeled as a $4.3\times 10^6\,M_{\odot}$ point mass sink.

\subsection{Tidal stripping and mass renormalization}

We first estimate the parameters of the tidal encounter. We analyze a scenario involving a brown dwarf (BD) with $M_{\text{BD}} = 0.01\,M_{\odot}$ orbiting Sgr A* ($M_{\text{BH}} = 4.3\times 10^6\,M_{\odot}$) with an illustrative periastron distance $R_p = 120\,R_{\odot}$. Assuming a standard BD radius $R_{\text{BD}} \approx 0.1\,R_{\odot}$, the tidal radius is
\begin{equation}
R_t \approx R_{\text{BD}} \left(\frac{M_{\text{BH}}}{M_{\text{BD}}}\right)^{1/3} \approx 75.5\,R_{\odot}.
\end{equation}
The penetration factor, $\beta = R_t/R_p \approx 0.63$, characterizes a grazing encounter ($\beta < 1$).

For such encounters, analytical models predict that a fraction of the stellar mass is tidally stripped. We estimate the stripped mass using the formalism of tidal energy deposition. The energy injected into the BD structure relative to its binding energy is given by
\begin{equation}
\frac{\Delta E_{\text{tide}}}{E_{\text{bind}}} \approx \left(\frac{M_{\text{BH}}}{M_{\text{BD}}}\right)^2 \left(\frac{R_{\text{BD}}}{R_p}\right)^6 T(\beta).
\end{equation}
Adopting a representative tidal coupling coefficient $T(\beta) \sim 10^{-4}$ appropriate for this penetration factor, we find $\Delta E_{\text{tide}}/E_{\text{bind}} \approx 6.3\times 10^{-6}$. For an $n=1.5$ polytropic structure appropriate for a BD, this yields an estimate of the analytically expected stripped mass:
\begin{equation}
\Delta M_{\text{analytical}} \approx {3.8\times 10^{-8}\,M_{\odot}.}
\label{eq:analytical_mass}
\end{equation}

We now examine the resolution constraints of our Smoothed Particle Hydrodynamics (SPH) simulations. We model the $0.01\,M_{\odot}$ BD using $N=250,000$ particles. This sets the mass resolution of the simulation, defined by the mass of a single SPH particle, $m_{\text{SPH}}$:
\begin{equation}
m_{\text{SPH}} = \frac{M_{\text{BD}}}{N} = 4.0\times 10^{-8}\,M_{\odot}.
\label{eq:sph_resolution}
\end{equation}

The critical constraint arises when considering the physical mass required to power the observed flare, given the parameters derived from our optimization workflow (see Sect.~\ref{sec:lightcurves}). To reproduce the observed data from Sgr A* with the derived non-thermal radiative efficiency $\eta_{\text{NT}}\approx 10^{-8}$, the required physical accreted mass is $M_{\text{physical}} \approx 1.5\times 10^{-7}\,M_{\odot}$.

Comparing this required physical mass to the resolution limit (Eq.~(\ref{eq:sph_resolution})), we find that $M_{\text{physical}} \approx 3.75 \cdot m_{\text{SPH}}$. A direct SPH simulation normalized to the physical BD mass would represent the entire accretion event with only approximately 4 particles. This number is insufficient to resolve the fluid dynamics of the stream.

The SPH methodology is necessary to capture the complex, three-dimensional gas dynamics of the tidal interaction. However, to resolve these dynamics accurately, we execute the SPH simulation using a high-mass proxy normalization. This proxy simulation provides the temporal evolution of the accretion rate, $\dot{M}_{\text{simulation}}(t)$, resolving the macrophysical dynamics of the interaction.

We must renormalize the simulation output in post-processing. We introduce a dimensionless rescaling factor, $E_{\text{strip}}$, to scale down the mass accretion rate from the simulation scale to the physical scale:
\begin{equation}
\dot{M}_{\text{physical}}(t) = E_{\text{strip}} \cdot \dot{M}_{\text{simulation}}(t).
\end{equation}
The SPH simulation provides the dynamical framework, while $E_{\text{strip}}$ adjusts the normalization to match the physical mass scale constrained by the observations and the required radiative efficiency. The derived conversion factor reconciles the simulation dynamics with the physical mass $M_{\text{physical}}$.

\section{Physical constraints on the non-thermal emission}
\label{sec:physics_constraints}

Our hydrodynamic simulations provide a robust estimate of the mass supply rate $\dot{M}(t)$ delivered to the vicinity of the black hole. However, the conversion of this mass flow into the observed non-thermal luminosity depends on complex plasma microphysics. In the previous sections, we encapsulated these uncertainties in the phenomenological efficiency parameter $\eta_{\text{NT}}$. Here, we critically examine the limitations of this approximation and provide a physical estimation of the emission mechanism based on synchrotron theory.

\subsection{Decoupling dynamics from microphysics}

The accretion event modeled here involves a hierarchy of physical timescales. The flare duration and light curve morphology are governed by the viscous time $t_{\nu}$ and the fallback time $t_{\text{fall}}$, both of which operate on macroscopic scales ($\sim 10^3 - 10^4$ s). Conversely, the particle acceleration and radiative cooling processes operate on kinetic timescales ($\ll 1$ s).

We explicitly state that we do not model the Magnetorotational Instability (MRI), magnetic reconnection, or relativistic jet launching in our SPH simulations. However, these effects operate on the local orbital timescale, which is much shorter than the global viscous timescale we simulate. Thus, they affect the efficiency (which we parameterize) but not the global dynamics (which we simulate). Consequently, while the \textit{amplitude} of the synthetic light curve depends on the assumed magnetic field amplification, the \textit{temporal profile}---the primary result of this work---is dynamically robust.

\subsection{Synchrotron formalism}

To validate the physical plausibility of the derived efficiency $\eta_{\text{NT}} \approx 10^{-8}$, we estimate the synchrotron luminosity expected from the simulated gas properties. We assume that the magnetic field in the accretion flow reaches equipartition with the dynamical pressure $P_{\text{dyn}}$ of the stream, such that $B^2 / 8\pi = \epsilon_B P_{\text{dyn}}$, where $\epsilon_B \lesssim 1$ is the equipartition parameter. From our SPH simulations near periapsis, the peak dynamical pressure typically reaches $P_{\text{dyn}} \sim 10^4$ dyne cm$^{-2}$, implying magnetic field strengths of order $B \sim 10^2$ G.

{The spectral power radiated by a single relativistic electron with Lorentz factor $\gamma$ and pitch angle $\alpha$ is governed by the classical Schwinger formalism involving the Airy function integral. The power per unit frequency is:}
\begin{equation}
    {P(\nu, \gamma, \alpha) = \frac{\sqrt{3} e^3 B \sin\alpha}{m_e c^2} F\left( \frac{\nu}{\nu_c} \right),}
    \label{eq:single_power}
\end{equation}
{where $\nu_c = \frac{3}{4\pi} \gamma^2 \frac{e B \sin\alpha}{m_e c}$ is the critical frequency. The spectral behavior is dictated by the dimensionless kernel $F(x)$, defined as the integral of the modified Bessel function of the second kind:}
\begin{equation}
    {F(x) = x \int_x^{\infty} K_{5/3}(\xi) d\xi.}
\end{equation}
{This function contains the microphysics: it yields the $\nu^{1/3}$ slope at low frequencies and the exponential cutoff at high frequencies.}

\subsection{Consistency check}

{The total non-thermal luminosity $L_{\text{NT}}$ is obtained by integrating Eq.~(\ref{eq:single_power}) over the frequency spectrum. Using the standard identity $\int_0^{\infty} F(x) dx = \frac{8\pi}{9\sqrt{3}}$, the angle-averaged total radiated power per electron corresponds to the standard result:}
\begin{equation}
    {P_{\text{avg}}(\gamma) = \frac{4}{3} \sigma_T c \gamma^2 U_B,}
\end{equation}
{where $U_B = B^2/8\pi$ is the magnetic energy density. For our estimated magnetic field $B \sim 100$ G and electrons accelerated to $\gamma \sim 10^3$ (consistent with the observed NIR spectral breaks), the cooling timescale is:}
\begin{equation}
    {t_{\text{cool}} = \frac{\gamma m_e c^2}{P_{\text{avg}}(\gamma)} \approx 77 \left( \frac{B}{100\,\text{G}} \right)^{-2} \left( \frac{\gamma}{10^3} \right)^{-1} \text{s}.}
\end{equation}
{This cooling time is significantly shorter than the dynamical flare duration ($\sim 3600$ s).} This confirms that the non-thermal emission is instantaneous with respect to the fuel supply, validating our assumption $L(t) \propto \dot{M}(t)$. Furthermore, to match the observed luminosity $L \approx 10^{36}$ erg s$^{-1}$, the required number density of relativistic electrons is a small fraction ($\sim 10^{-4}$) of the total gas density derived from the simulations. This confirms that the low global efficiency $\eta_{\text{NT}} \sim 10^{-8}$ is physically robust, corresponding to a scenario where only a trace population of particles is accelerated within a bulk flow that remains largely non-radiative.

\begin{figure}[t]
   \centering
   \includegraphics[width=0.45\textwidth]{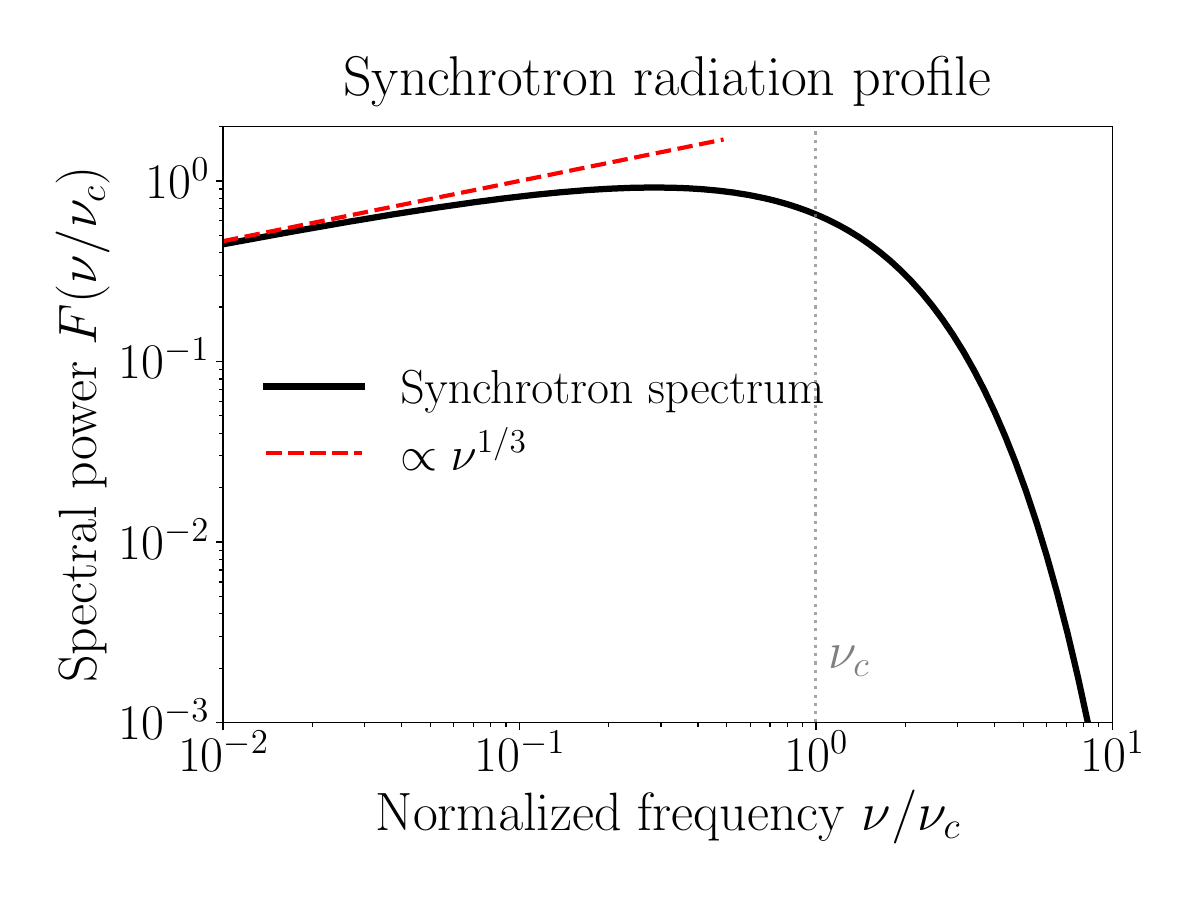}
   \caption{
   The universal synchrotron spectral power function $F(x)$ as a function of the normalized frequency $x \equiv \nu/\nu_c$.
   This profile dictates the spectral energy distribution of the relativistic electron population assumed in our model.
   The function exhibits a distinct maximum at $x \approx 0.29$, reaching a peak amplitude of $F(x) \approx 0.92$.
   In the low-frequency regime ($x \ll 1$), the emission rises according to the classical asymptotic power law $F(x) \propto x^{1/3}$.
   The high-frequency domain ($x > 1$) demonstrates a rapid exponential suppression; the spectral power contracts to $F(1) \approx 0.65$ at the critical frequency and diminishes by four orders of magnitude by $x \approx 10$ ($F(10) \sim 10^{-4}$).
   This characteristic cutoff defines the cooling break in the spectral energy distribution, ensuring that the modeled X-ray flux remains consistent with the observational upper limits while permitting bright non-thermal emission in the infrared.
   }
   \label{fig:airy}
\end{figure}

To validate the physical plausibility of the assumed radiative efficiency, we compute the spectral signature of the emitting plasma.
Figure~\ref{fig:airy} illustrates the spectral power distribution for a single relativistic electron, derived from the Airy integral formalism.
The profile reveals the two fundamental regimes governing the emission: the optically thin power-law ascent at low frequencies and the precipitous exponential decay beyond the critical frequency $\nu_c$.
The location of the spectral peak at $\nu \approx 0.29 \nu_c$ serves as a robust diagnostic for the electron energy distribution; for the magnetic field strengths of $B \sim 10^2$ G estimated from the dynamical pressure of the simulated stream, this peak aligns with the near-infrared band for Lorentz factors of $\gamma \sim 10^3$.
Furthermore, the steep exponential cutoff at high frequencies provides a natural physical mechanism for the observed spectral softening between the infrared and X-ray bands, confirming that the cooling break occurs at energies consistent with the lack of persistent bright X-ray emission during the quiescent phases.
Consequently, the rapid radiative cooling implied by the integral of this spectral form justifies the assumption that the non-thermal luminosity instantaneously tracks the mass accretion rate.

\end{appendix}	


\begin{thebibliography}{24}
\expandafter\ifx\csname natexlab\endcsname\relax\def\natexlab#1{#1}\fi

\bibitem[{{Amaro-Seoane}(2019)}]{AmaroSeoane2019}
{Amaro-Seoane}, P. 2019, \prd, 99, 123025

\bibitem[{{Amaro Seoane} \& {Zhao}(2025)}]{AmaroSeoaneShaoDong2025}
{Amaro Seoane}, P. \& {Zhao}, S.-D. 2025, arXiv e-prints, arXiv:2504.20147

\bibitem[{{Calder{\'o}n} {et~al.}(2025){Calder{\'o}n}, {Cuadra}, {Russell},
  {Burkert}, {Rosswog}, \& {Balakrishnan}}]{CalderonEtAl2025}
{Calder{\'o}n}, D., {Cuadra}, J., {Russell}, C. M.~P., {et~al.} 2025, \aap,
  693, A180

\bibitem[{{Dodds-Eden} {et~al.}(2009){Dodds-Eden}, {Porquet}, {Trap},
  {Quataert}, {Haubois}, {Gillessen}, {Grosso}, {Pantin}, {Falcke}, {Rouan},
  {Genzel}, {Hasinger}, {Goldwurm}, {Yusef-Zadeh}, {Clenet}, {Trippe},
  {Lagage}, {Bartko}, {Eisenhauer}, {Ott}, {Paumard}, {Perrin}, {Yuan},
  {Fritz}, \& {Mascetti}}]{DoddsEdenEtAl2009}
{Dodds-Eden}, K., {Porquet}, D., {Trap}, G., {et~al.} 2009, \apj, 698, 676

\bibitem[{{Eggleton}(1983)}]{Eggleton1983}
{Eggleton}, P.~P. 1983, \apj, 268, 368

\bibitem[{{Gravity Collaboration} {et~al.}(2023){Gravity Collaboration},
  {Abuter}, {Aimar}, {Amaro Seoane}, {Amorim}, {Baub{\"o}ck}, {Berger},
  {Bonnet}, {Bourdarot}, {Brandner}, {Cardoso}, {Cl{\'e}net}, {Davies}, {de
  Zeeuw}, {Dexter}, {Drescher}, {Eckart}, {Eisenhauer}, {Feuchtgruber},
  {Finger}, {F{\"o}rster Schreiber}, {Foschi}, {Garcia}, {Gao}, {Gelles},
  {Gendron}, {Genzel}, {Gillessen}, {Hartl}, {Haubois}, {Haussmann},
  {Hei{\ss}el}, {Henning}, {Hippler}, {Horrobin}, {Jochum}, {Jocou}, {Kaufer},
  {Kervella}, {Lacour}, {Lapeyr{\`e}re}, {Le Bouquin}, {L{\'e}na}, {Lutz},
  {Mang}, {More}, {Ott}, {Paumard}, {Perraut}, {Perrin}, {Pfuhl}, {Rabien},
  {Ribeiro}, {Sadun Bordoni}, {Scheithauer}, {Shangguan}, {Shimizu}, {Stadler},
  {Straub}, {Straubmeier}, {Sturm}, {Tacconi}, {Vincent}, {von Fellenberg},
  {Widmann}, {Wielgus}, {Wieprecht}, {Wiezorrek}, \& {Woillez}}]{GRAVITY2023}
{Gravity Collaboration}, {Abuter}, R., {Aimar}, N., {et~al.} 2023, \aap, 677,
  L10

\bibitem[{{GRAVITY Collaboration} {et~al.}(2021){GRAVITY Collaboration},
  {Abuter}, {Amorim}, {Baub{\"o}ck}, {Baganoff}, {Berger}, {Boyce}, {Bonnet},
  {Brandner}, {Cl{\'e}net}, {Davies}, {de Zeeuw}, {Dexter}, {Dallilar},
  {Drescher}, {Eckart}, {Eisenhauer}, {Fazio}, {F{\"o}rster Schreiber},
  {Foster}, {Gammie}, {Garcia}, {Gao}, {Gendron}, {Genzel}, {Ghisellini},
  {Gillessen}, {Gurwell}, {Habibi}, {Haggard}, {Hailey}, {Harrison}, {Haubois},
  {Hei{\ss}el}, {Henning}, {Hippler}, {Hora}, {Horrobin},
  {Jim{\'e}nez-Rosales}, {Jochum}, {Jocou}, {Kaufer}, {Kervella}, {Lacour},
  {Lapeyr{\`e}re}, {Le Bouquin}, {L{\'e}na}, {Lowrance}, {Lutz}, {Markoff},
  {Mori}, {Morris}, {Neilsen}, {Nowak}, {Ott}, {Paumard}, {Perraut}, {Perrin},
  {Ponti}, {Pfuhl}, {Rabien}, {Rodr{\'\i}guez-Coira}, {Shangguan}, {Shimizu},
  {Scheithauer}, {Smith}, {Stadler}, {Stern}, {Straub}, {Straubmeier}, {Sturm},
  {Tacconi}, {Vincent}, {von Fellenberg}, {Waisberg}, {Widmann}, {Wieprecht},
  {Wiezorrek}, {Willner}, {Witzel}, {Woillez}, {Yazici}, {Young}, {Zhang}, \&
  {Zins}}]{GRAVITY2021}
{GRAVITY Collaboration}, {Abuter}, R., {Amorim}, A., {et~al.} 2021, \aap, 654,
  A22

\bibitem[{{GRAVITY Collaboration} {et~al.}(2020){GRAVITY Collaboration},
  {Abuter}, {Amorim}, {Baub{\"o}ck}, {Berger}, {Bonnet}, {Brandner}, {Cardoso},
  {Cl{\'e}net}, {de Zeeuw}, {Dallilar}, {Dexter}, {Eckart}, {Eisenhauer},
  {F{\"o}rster Schreiber}, {Garcia}, {Gao}, {Gendron}, {Genzel}, {Gillessen},
  {Habibi}, {Haubois}, {Henning}, {Hippler}, {Horrobin}, {Jim{\'e}nez-Rosales},
  {Jochum}, {Jocou}, {Kaufer}, {Kervella}, {Lacour}, {Lapeyr{\`e}re}, {Le
  Bouquin}, {L{\'e}na}, {Nowak}, {Ott}, {Paumard}, {Perraut}, {Perrin},
  {Pfuhl}, {Ponti}, {Rodriguez Coira}, {Shangguan}, {Scheithauer}, {Stadler},
  {Straub}, {Straubmeier}, {Sturm}, {Tacconi}, {Vincent}, {von Fellenberg},
  {Waisberg}, {Widmann}, {Wieprecht}, {Wiezorrek}, {Woillez}, {Yazici}, \&
  {Zins}}]{GRAVITY_Flux2020}
{GRAVITY Collaboration}, {Abuter}, R., {Amorim}, A., {et~al.} 2020, \aap, 638,
  A2

\bibitem[{{Lombardi}(1998)}]{Lombardi1998}
{Lombardi}, J.~C. 1998, PhD thesis, Cornell University, New York

\bibitem[{{Lombardi} {et~al.}(1999){Lombardi}, {Sills}, {Rasio}, \&
  {Shapiro}}]{LombardiEtAl1999}
{Lombardi}, J.~C., {Sills}, A., {Rasio}, F.~A., \& {Shapiro}, S.~L. 1999,
  Journal of Computational Physics, 152, 687

\bibitem[{{Paxton} {et~al.}(2011){Paxton}, {Bildsten}, {Dotter}, {Herwig},
  {Lesaffre}, \& {Timmes}}]{Paxton2011}
{Paxton}, B., {Bildsten}, L., {Dotter}, A., {et~al.} 2011, \apjs, 192, 3

\bibitem[{{Paxton} {et~al.}(2013){Paxton}, {Cantiello}, {Arras}, {Bildsten},
  {Brown}, {Dotter}, {Mankovich}, {Montgomery}, {Stello}, {Timmes}, \&
  {Townsend}}]{Paxton2013}
{Paxton}, B., {Cantiello}, M., {Arras}, P., {et~al.} 2013, \apjs, 208, 4

\bibitem[{{Pfuhl} {et~al.}(2015){Pfuhl}, {Gillessen}, {Eisenhauer}, {Genzel},
  {Plewa}, {Ott}, {Ballone}, {Schartmann}, {Burkert}, {Fritz}, {Sari},
  {Steinberg}, \& {Madigan}}]{PfuhlEtAl2015}
{Pfuhl}, O., {Gillessen}, S., {Eisenhauer}, F., {et~al.} 2015, \apj, 798, 111

\bibitem[{{Ponti} {et~al.}(2015){Ponti}, {De Marco}, {Morris}, {Merloni},
  {Mu{\~n}oz-Darias}, {Clavel}, {Haggard}, {Zhang}, {Nandra}, {Gillessen},
  {Mori}, {Neilsen}, {Rea}, {Degenaar}, {Terrier}, \&
  {Goldwurm}}]{PontiEtAl2015}
{Ponti}, G., {De Marco}, B., {Morris}, M.~R., {et~al.} 2015, \mnras, 454, 1525

\bibitem[{{Rasio}(1991)}]{Rasio1991}
{Rasio}, F.~A. 1991, PhD thesis, Cornell University, New York

\bibitem[{{Rasio} \& {Lombardi}(1999)}]{RasioLombardi1999}
{Rasio}, F.~A. \& {Lombardi}, Jr., J.~C. 1999, Journal of Computational and
  Applied Mathematics, 109, 213

\bibitem[{{Ressler} {et~al.}(2019){Ressler}, {Quataert}, \&
  {Stone}}]{ResslerEtAl2019}
{Ressler}, S.~M., {Quataert}, E., \& {Stone}, J.~M. 2019, \mnras, 482, L123

\bibitem[{{Timmer} \& {K{\"o}nig}(1995)}]{TimmerKoenig1995}
{Timmer}, J. \& {K{\"o}nig}, M. 1995, \aap, 300, 707

\bibitem[{{V{\'a}zquez-Aceves} {et~al.}(2023){V{\'a}zquez-Aceves}, {Lin}, \&
  {Torres-Orjuela}}]{VazquezAcevesEtAl2023}
{V{\'a}zquez-Aceves}, V., {Lin}, Y., \& {Torres-Orjuela}, A. 2023, \apj, 952,
  139

\bibitem[{{V{\'a}zquez-Aceves} {et~al.}(2025){V{\'a}zquez-Aceves}, {Lin}, \&
  {Torres-Orjuela}}]{VazquezAcevesEtAl2025}
{V{\'a}zquez-Aceves}, V., {Lin}, Y., \& {Torres-Orjuela}, A. 2025, \apj, 987,
  208

\bibitem[{{von Fellenberg} {et~al.}(2025){von Fellenberg}, {Roychowdhury},
  {Michail}, {Sumners}, {Sanger-Johnson}, {Fazio}, {Haggard}, {Hora},
  {Philippov}, {Ripperda}, {Smith}, {Willner}, {Witzel}, {Zhang}, {Becklin},
  {Bower}, {Chandra}, {Do}, {Garcia Marin}, {Gurwell}, {Ford}, {Hada},
  {Markoff}, {Morris}, {Neilsen}, {Sabha}, \& {Seefeldt-Gail}}]{GRAVITY2025}
{von Fellenberg}, S.~D., {Roychowdhury}, T., {Michail}, J.~M., {et~al.} 2025,
  \apjl, 979, L20

\bibitem[{{Witzel} {et~al.}(2014){Witzel}, {Ghez}, {Morris}, {Sitarski},
  {Boehle}, {Naoz}, {Campbell}, {Becklin}, {Canalizo}, {Chappell}, {Do}, {Lu},
  {Matthews}, {Meyer}, {Stockton}, {Wizinowich}, \& {Yelda}}]{WitzelEtAl2014}
{Witzel}, G., {Ghez}, A.~M., {Morris}, M.~R., {et~al.} 2014, \apjl, 796, L8

\bibitem[{{Witzel} {et~al.}(2018){Witzel}, {Martinez}, {Hora}, {Willner},
  {Morris}, {Gammie}, {Becklin}, {Ashby}, {Baganoff}, {Carey}, {Do}, {Fazio},
  {Ghez}, {Glaccum}, {Haggard}, {Herrero-Illana}, {Ingalls}, {Narayan}, \&
  {Smith}}]{WitzelEtAl2018A}
{Witzel}, G., {Martinez}, G., {Hora}, J., {et~al.} 2018, \apj, 863, 15

\bibitem[{{Yuan} {et~al.}(2003){Yuan}, {Quataert}, \& {Narayan}}]{YuanEtAl2003}
{Yuan}, F., {Quataert}, E., \& {Narayan}, R. 2003, \apj, 598, 301

\end{thebibliography}
\end{document}